\documentclass[showpacs,aps,graphicx,onecolumn,10pt]{revtex4-2}
\usepackage{amsmath}
\usepackage{mathrsfs}
\usepackage{amsfonts}
\usepackage{amssymb}
\usepackage{graphicx}
\usepackage{caption}
\usepackage{subfigure}
\usepackage{eufrak}
\usepackage{multirow}
\usepackage{float}
\usepackage[colorlinks,linkcolor=blue,anchorcolor=blue,citecolor=blue]{hyperref}
\usepackage{soul,color,xcolor}

\begin{document}

\title{Heralded high-dimensional module-based quantum computation}

\author{Xiao Zhang\textsuperscript{1}, Wen-Qiang Liu\textsuperscript{2} and Hai-Rui Wei\textsuperscript{1,}}
\email[]{hrwei@ustb.edu.cn}
\address{\textsuperscript{\rm1} School of Mathematics and Physics, University of Science and Technology Beijing, Beijing 100083, China\\
\textsuperscript{\rm2} Department of Mathematics and Physics, Shijiazhuang Tiedao University, Shijiazhuang 050043, China}

\begin{abstract}
Parity measurements have been explored as building blocks for preparing and discriminating entangled states, as well as for implementing quantum computation.  
We first develop two alternative high-dimensional generalized parity modules, and then propose a procedure for constructing high-dimensional generalized module-based controlled-NOT gate.
The construction of module-based quantum computing introduced here is deterministic, heralded, insensitive to the dimensionality of the computing basis, and postselection technique is not required. The result shows that out of $(d-1)!$ generalized parity modules, only modules $ \mathcal{P}=(j\ominus i)\bmod d$ and  $ \mathcal{P}=(j \oplus i)\bmod d$ can be used as building blocks for high-dimensional quantum computing.
Furthermore, we proposed an optical nondestructive scheme for implementing generalized parity module through quantum nondemolition measurements, and the success of the parity module is heralded by photon-number-resolving detectors and single-photon detectors.
\end{abstract}


\maketitle

\section{Introduction} \label{sec1}

Quantum computation can outperform its classical counterparts in efficiently solving certain hard computational problems \cite{book}. 
Quantum gates are essential for realizing quantum computation,  hence it is significant to find a simple and efficient way to implement them. 
Qubit-based controlled-NOT (CNOT) gates are the most widely used universal gates because CNOTs assisted by single-qubit gates are sufficient to implement any multiqubit unitary quantum computation \cite{Barenco}.
The theoretical lower bound for the complexity of an $n$-qubit quantum computation, which is usually measured by assessing the number of CNOT gates required to perform computation, is $[(4^n-3n-1)/4]$ CNOT gates \cite{lower-bound}. 
The implementations of Toffoli \cite{Toffoli} and Fredkin gates \cite{Fredkin} are usually investigated using the standard CNOT-based approach.
CNOT gates also play an important role in certain quantum communication tasks, such as entangled state generation, and entanglement purification \cite{high-generation}. 
To date, there have been many theoretical proposals \cite{CNOT-Theo, Han2021, Su2024} and realistic experiments \cite{CNOT-Exp} for realizing CNOT gates, which have now been demonstrated across diverse physical platforms, including photons \cite{CNOT-photon}, trapped ions \cite{CNOT-ion}, superconductors \cite{CNOT-superconductor}, diamond nitrogen-vacancy centers \cite{CNOT-NV}, and atoms \cite{CNOT-atom}.

Compared with binary quantum systems, high-dimensional quantum systems, so called qudits ($d$-level or $d$-state systems with $d > 2$), offer substantial advantages, such as increased channel capacity \cite{capacity1,capacity2}, enhanced noise resilience \cite{resilience1,resilience2, Vinet2025}, reduced algorithmic complexity \cite{reduced-complexity1,reduced-complexity2}, improved efficiency and accuracy \cite{efficiency-accurary, Gong2024}, simplified experimental setups \cite{simplified}, and strengthened violation of Bell's non-locality \cite{non-locality}. 
Although high-dimensional quantum systems exhibit remarkable advantages in various applications and substantial potential for future development, particularly in high-dimensional Bell state measurements and superdense coding \cite{high, OL2022BSM, JOSAB2025superdense, JOSAB2025interference}, these systems have been less thoroughly investigated, both theoretically and experimentally, than conventional two-dimensional systems. 
Nowadays, various physical platforms have been proposed to implement high-dimensional quantum computing, including photons (using $d$ optical modes in path, frequency, orbital angular momentum (OAM), and time-bin degrees of freedom (DOFs)) \cite{high-photon, Shafran2026}, ion traps \cite{high-ion}, continuous spin systems \cite{high-spin}, diamond nitrogen-vacancy centers \cite{high-NV}, nuclear magnetic resonances \cite{high-NMR}, molecular magnets \cite{high-mmolecular}, and superconducting devices \cite{high-supercond}.

In the field of single-qudit gates, generalized Pauli (generalized $X_d$ and $Z_d$) gates in OAM-based optical qudits \cite{single-qudit-Pauli} and Walsh-Hadamard (Fourier) gates in superconducting qutrits (i.e., 3-level or 3-state systems) \cite{single-qudit-Hadamard} have been experimentally demonstrated. 
Moreover, it has been demonstrated that arbitrary single-qudit spatial-based gates can be well manipulated with high precision using programmable interferometers \cite{interferometer1, interferometer2}. 
Recently, high-precision universal single-qudit operations have been experimentally demonstrated on trapped-ion platforms up to $d=7$ \cite{simplified} and superconducting qutrit ($d=3$) devices with an average fidelity of $3.8 \times 10^{-3}$ \cite{Morvan2021qutrit}.
In the field of two-qudit universal gates, various generalized CNOT-type gates \cite{high-Kerr, Duqudit}, such as controlled-SUM, controlled-increment (CINC), controlled-XOR (GXOR), and controlled-$X$ (GCX) gates, have been proposed to synthesize general qudit circuits. 
Based on quantum Shannon decomposition, Di and Wei \cite{Di-Wei} showed that the best-known lower bound arbitrary $n$-qudit gate is $[d^{2n}-n(d^2-1)-1]/[4(d-1)]$ GCX. 
In 2009, Nakajima \emph{et al.} \cite{Nakajima} demonstrated that the best-known cost of general $n$-qudit quantum computing is $O(n^{2+\log_2 d} d^{2n})$ CINCs when $d$ is odd, $O(d^{2n})$ CINCs when $d$ is a power of two, and $O(n d^{2n})$ CINCs otherwise.
Recently, by refining Cartan decomposition technique, Jiang and Wei \cite{Jiang-Wei} have presented the best-known circuits for general two-qutrit and quNit-quMit quantum computation in terms of CINCs and GCXs.
However, architectures for the above qudit universal gates are rarely investigated both theoretically and experimentally.   

In this paper, we propose schemes to construct deterministic generalized qudit-based CNOT gates using generalized parity modules.  
We first propose various high-dimensional generalized parity measurements. 
Then, utilizing the proposed building blocks, the high-dimensional generalized CNOT gates are constructed.  
Subsequently, we design an optical architecture for implementing the OAM-based generalized parity module.  
The proposed program exhibits the following characteristics: 
(i) The protocols are deterministic and the computational qudits are not consumed in principle. 
(ii) Large-scale entangled states are not required, and only one additional qudit is employed in the proposed scheme. 
(iii) The success of the proposed high-dimensional optical nondestructive generalized parity module operation can be heralded by photon-number-resolving detectors and single-photon detectors. 
Moreover, the proposed generalized parity module can also be applied to quantum communication.


\section{Module-based generalized two-qutrit CNOT gate} \label{Sec2}

\begin{figure}
\centering
\includegraphics[width=8.5 cm]{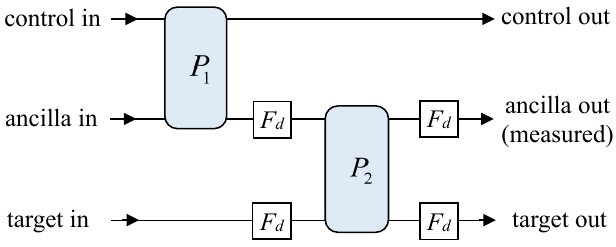}
\caption{The schematic diagram for constructing a deterministic high-dimensional generalized CNOT gate.
         Blocks $P_1$ and $P_2$ are two generalized parity modules.
         $F_d$ denotes the qudit-based Fourier transformation, which performs the transformation 
         $|m\rangle\rightarrow\frac{1}{\sqrt{d}} \sum_{n=0}^{d-1} e^{\mathrm{i} 2\pi m n / d} |n\rangle$ with $ m=0,1,\cdots, (d-1)$. 
         The ancilla qudit is initially prepared in the state $\frac{1}{\sqrt{d}}(|0\rangle + |1\rangle + \cdots +|d-1\rangle)$.}
\label{Fig1}
\end{figure}

In this section, we develop a scheme for constructing a two-qutrit generalized CNOT gate via generalized parity modules. 
The operation of the two-qudit generalized CNOT gate is given by
\begin{equation} \label{eq1}
\mathrm{CNOT}|x\rangle_c |y\rangle_t = |x\rangle_c |(x\oplus y) \bmod d\rangle_t,
\end{equation}
where $x,y \in \{0,1,\cdots,(d-1)\}$.

The qudit-based generalized parity module, also referred to as a generalized quantum gate, has two inputs $|i\rangle$ and $|j\rangle$ and an ancilla $|0\rangle$, $i,j \in \{0,1, \cdots (d-1)\}$.  
The action of an alternative generalized parity module is defined as follows 
\begin{equation} \label{eq2}
|i j \rangle |0\rangle \rightarrow |i j\rangle |(j \ominus i) \bmod d\rangle.
\end{equation}
That is, it leaves the basis states $|i\rangle$ and $|j\rangle$ invariant, whereas it outputs $ \mathcal{P} =(j \ominus i) \bmod d$.   
  
\begin{table} [tbp]
\centering
\caption{The correspondences between the basis and the outputs in qutrit-based generalized parity module with $\mathcal{P} =(j \ominus i) \bmod 3$. 
         Here $|i\rangle$ and $|j\rangle$ are two input states.}
\label{Tab1}

\begin{tabular}{cccccccccc}
\hline\hline

\multicolumn {3}{c}{Basis}                    &  Output $\mathcal{P}$     & Class \\

\hline

 $|00\rangle$ & $|11\rangle$  & $|22\rangle$  &   0                       & \uppercase\expandafter{\romannumeral1} \\

 $|01\rangle$ & $|12\rangle$  & $|20\rangle$  &   1                       & \uppercase\expandafter{\romannumeral2} \\

 $|02\rangle$ & $|10\rangle$  & $|21\rangle$  &   2                       & \uppercase\expandafter{\romannumeral3} \\
 
\hline \hline
\end{tabular}
\end{table}

Now, let us explain the detailed procedure of our two-qutrit generalized CNOT gate via generalized parity modules, step by step.
Suppose that the control and the target qutrits are initially prepared in the arbitrary states
\begin{equation} \label{eq3}
|x\rangle_c = \alpha_0|0\rangle_c + \alpha_1|1\rangle_c + \alpha_2|2\rangle_c, \;\;
|y\rangle_t = \beta_0|0\rangle_t  + \beta_1 |1\rangle_t + \beta_2 |2\rangle_t,
\end{equation}
%
%
where the complex coefficients $\alpha_0$, $\alpha_1$, $\alpha_2$, $\beta_0$, $\beta_1$, and $\beta_2$ satisfy 
                               $|\alpha_0|^2 + |\alpha_1|^2 + |\alpha_2|^2 = 1$ and $|\beta_0|^2 + |\beta_1|^2 + |\beta_2|^2 = 1$.
The ancilla qutrit is initially prepared in the balanced state
\begin{equation} \label{eq4}
|z\rangle_a = \frac{1}{\sqrt{3}}(|0\rangle_a +|1\rangle_a +|2\rangle_a).
\end{equation}
Consequently, the initial state of the composite system composed of control, target, and ancilla qutrits is expressed as 
\begin{equation} \label{eq5}
|\psi_\text{in}\rangle = |x\rangle_c \otimes |y\rangle_t \otimes |z\rangle_a.
\end{equation}

In the first step,  the control and ancilla qutrits are injected into block $P_1$. 
As shown in Table \ref{Tab1}, 
if the output of block $P_1$ is $\mathcal{P}_1=0$, then $|\psi_\text{in}\rangle$ collapses to 
\begin{equation} \label{eq6}
|\psi_0\rangle = (\alpha_0|0\rangle_c|0\rangle_a 
                + \alpha_1|1\rangle_c|1\rangle_a 
                + \alpha_2|2\rangle_c|2\rangle_a) \otimes |y\rangle_t.
\end{equation}
%
If $\mathcal{P}_1=1$, then $|\psi_\text{in}\rangle$ collapses to 
\begin{equation} \label{eq7}
|\psi_1\rangle = (\alpha_0|0\rangle_c|1\rangle_a 
                + \alpha_1|1\rangle_c|2\rangle_a 
                + \alpha_2|2\rangle_c|0\rangle_a) \otimes |y\rangle_t.
\end{equation}
%
If $\mathcal{P}_1=2$, then $|\psi_\text{in}\rangle$ collapses to 
\begin{equation} \label{eq8}
|\psi_2\rangle = (\alpha_0|0\rangle_c|2\rangle_a 
                + \alpha_1|1\rangle_c|0\rangle_a 
                + \alpha_2|2\rangle_c|1\rangle_a) \otimes |y\rangle_t.
\end{equation}

In the second step, as depicted in Fig. \ref{Fig1}, block $P_2$ is performed on the ancilla and the target qutrits. 
Before and after $P_2$, single-qutrit Fourier transformations, $F_3$s, are performed on the ancilla and target qutrits, respectively.
Here, the matrix representation of $F_3$ is given by
\begin{equation} \label{eq9}
F_{3}=\frac{1}{\sqrt{3}}\left(
                          \begin{array}{ccc}
                            1 & 1                            & 1 \\
                            1 & e^{\text{i} \frac{2\pi}{3}}  & e^{\text{i} \frac{4\pi}{3}} \\
                            1 & e^{\text{i} \frac{4\pi}{3}}  & e^{\text{i} \frac{2\pi}{3}}
                          \end{array}
                        \right)
\end{equation} 
in the basis $\{|0\rangle, |1\rangle, |2\rangle\}$.
Based on Table \ref{Tab1} and Eq. \eqref{eq9}, one can find that after the above operations ($F_3 \rightarrow P_2 \rightarrow F_3$),
%
if $\mathcal{P}_2=0$, then state $|\psi_0\rangle$ becomes
\begin{equation} \label{eq10}
\begin{split}
|\psi_{00}\rangle = &\ 
    \alpha_0|0\rangle_c[\beta_0(|00\rangle + |12\rangle + |21\rangle)_{ta}    
                      + \beta_1(|02\rangle + |11\rangle + |20\rangle)_{ta}\\& 
                      + \beta_2(|01\rangle + |10\rangle + |22\rangle)_{ta}]
  + \alpha_1|1\rangle_c[\beta_0(|02\rangle + |11\rangle + |20\rangle)_{ta}\\& 
                      + \beta_1(|01\rangle + |10\rangle + |22\rangle)_{ta}    
                      + \beta_2(|00\rangle + |12\rangle + |21\rangle)_{ta}]\\&
  + \alpha_2|2\rangle_c[\beta_0(|01\rangle + |10\rangle + |22\rangle)_{ta}    
                      + \beta_1(|00\rangle + |12\rangle + |21\rangle)_{ta}\\& 
                      + \beta_2(|02\rangle + |11\rangle + |20\rangle)_{ta}];
\end{split}
\end{equation}
State $|\psi_1\rangle$ becomes
\begin{equation} \label{eq11}
\begin{split}
|\psi_{10}\rangle = &\ 
    \alpha_0|0\rangle_c[\beta_0(|02\rangle + |11\rangle + |20\rangle)_{ta}    
                      + \beta_1(|01\rangle + |10\rangle + |22\rangle)_{ta}\\& 
                      + \beta_2(|00\rangle + |12\rangle + |21\rangle)_{ta}]
  + \alpha_1|1\rangle_c[\beta_0(|01\rangle + |10\rangle + |22\rangle)_{ta}\\& 
                      + \beta_1(|00\rangle + |12\rangle + |21\rangle)_{ta}    
                      + \beta_2(|02\rangle + |11\rangle + |20\rangle)_{ta}]\\&
  + \alpha_2|2\rangle_c[\beta_0(|00\rangle + |12\rangle + |21\rangle)_{ta}    
                      + \beta_1(|02\rangle + |11\rangle + |20\rangle)_{ta}\\& 
                      + \beta_2(|01\rangle + |10\rangle + |22\rangle)_{ta}];
\end{split}
\end{equation}
State $|\psi_2\rangle$ becomes 
\begin{equation} \label{eq12}
\begin{split}
|\psi_{20}\rangle  =&\ 
    \alpha_0|0\rangle_c[\beta_0(|01\rangle + |10\rangle + |22\rangle)_{ta} 
                      + \beta_1(|00\rangle + |12\rangle + |21\rangle)_{ta}\\&
                      + \beta_2(|02\rangle + |11\rangle + |20\rangle)_{ta}]
  + \alpha_1|1\rangle_c[\beta_0(|00\rangle + |12\rangle + |21\rangle)_{ta} \\&
                      + \beta_1(|02\rangle + |11\rangle + |20\rangle)_{ta}
                      + \beta_2(|01\rangle + |10\rangle + |22\rangle)_{ta}]\\&
  + \alpha_2|2\rangle_c[\beta_0(|02\rangle + |11\rangle + |20\rangle)_{ta} 
                      + \beta_1(|01\rangle + |10\rangle + |22\rangle)_{ta}\\&
                      + \beta_2(|00\rangle + |12\rangle + |21\rangle)_{ta}].
\end{split}
\end{equation}

If $\mathcal{P}_2=1$, then state $|\psi_0\rangle$ becomes 
\begin{equation} \label{eq13}
\begin{split}
|\psi_{01}\rangle =&\ 
    \alpha_0|0\rangle_c[\beta_0(|00\rangle                            + e^{\text{i} \frac{4\pi}{3}}|12\rangle           + e^{\text{i} \frac{2\pi}{3}}|21\rangle)_{ta} 
                      + \beta_1(e^{\text{i} \frac{4\pi}{3}}|02\rangle + e^{\text{i} \frac{2\pi}{3}}|11\rangle           +                            |20\rangle)_{ta}\\&
                      + \beta_2(e^{\text{i} \frac{2\pi}{3}}|01\rangle +                            |10\rangle           + e^{\text{i} \frac{4\pi}{3}}|22\rangle)_{ta}]
  + \alpha_1|1\rangle_c[\beta_0(|02\rangle                            + e^{\text{i}\frac{4\pi}{3}} |11\rangle           + e^{\text{i}\frac{2\pi}{3}} |20\rangle)_{ta} \\&
                      + \beta_1(e^{\text{i} \frac{4\pi}{3}}|01\rangle + e^{\text{i} \frac{2\pi}{3}}|10\rangle           +                            |22\rangle)_{ta}
                      + \beta_2(e^{\text{i} \frac{2\pi}{3}}|00\rangle +                            |12\rangle           + e^{\text{i} \frac{4\pi}{3}}|21\rangle)_{ta}]\\&
  + \alpha_2|2\rangle_c[\beta_0(|01\rangle                            + e^{\text{i} \frac{4\pi}{3}}|10\rangle           + e^{\text{i} \frac{2\pi}{3}}|22\rangle)_{ta} 
                      + \beta_1(e^{\text{i} \frac{4\pi}{3}}|00\rangle + e^{\text{i} \frac{2\pi}{3}}|12\rangle           +                            |21\rangle)_{ta}\\&
                      + \beta_2(e^{\text{i} \frac{2\pi}{3}}|02\rangle +                            |11\rangle           + e^{\text{i} \frac{4\pi}{3}}|20\rangle)_{ta}];
\end{split}
\end{equation}
State $|\psi_1\rangle$ becomes
\begin{equation} \label{eq14}
\begin{split}
|\psi_{11}\rangle =&\ 
    \alpha_0|0\rangle_c[\beta_0(                           |02\rangle + e^{\text{i} \frac{4\pi}{3}}|11\rangle  + e^{\text{i} \frac{2\pi}{3}} |20\rangle)_{ta} 
                      + \beta_1(e^{\text{i} \frac{4\pi}{3}}|01\rangle + e^{\text{i} \frac{2\pi}{3}}|10\rangle  +                             |22\rangle)_{ta}\\&
                      + \beta_2(e^{\text{i} \frac{2\pi}{3}}|00\rangle +                            |12\rangle  + e^{\text{i} \frac{4\pi}{3}} |21\rangle)_{ta}]
  + \alpha_1|1\rangle_c[\beta_0(                           |01\rangle + e^{\text{i} \frac{4\pi}{3}}|10\rangle  + e^{\text{i} \frac{2\pi}{3}} |22\rangle)_{ta} \\&
                      + \beta_1(e^{\text{i} \frac{4\pi}{3}}|00\rangle + e^{\text{i} \frac{2\pi}{3}}|12\rangle  +                             |21\rangle)_{ta}
                      + \beta_2(e^{\text{i} \frac{2\pi}{3}}|02\rangle +                            |11\rangle  + e^{\text{i} \frac{4\pi}{3}} |20\rangle)_{ta}]\\&
  + \alpha_2|2\rangle_c[\beta_0(                           |00\rangle + e^{\text{i} \frac{4\pi}{3}}|12\rangle  + e^{\text{i} \frac{2\pi}{3}} |21\rangle)_{ta} 
                      + \beta_1(e^{\text{i} \frac{4\pi}{3}}|02\rangle + e^{\text{i} \frac{2\pi}{3}}|11\rangle  +                             |20\rangle)_{ta}\\&
                      + \beta_2(e^{\text{i} \frac{2\pi}{3}}|01\rangle +                            |10\rangle  + e^{\text{i} \frac{4\pi}{3}} |22\rangle)_{ta}];
\end{split}
\end{equation}
State $|\psi_2\rangle$ becomes 
\begin{equation} \label{eq15}
\begin{split}
|\psi_{21}\rangle =&\ 
    \alpha_0|0\rangle_c[\beta_0(                           |01\rangle + e^{\text{i} \frac{4\pi}{3}}|10\rangle +  e^{\text{i} \frac{2\pi}{3}}|22\rangle)_{ta} 
                      + \beta_1(e^{\text{i} \frac{4\pi}{3}}|00\rangle + e^{\text{i} \frac{2\pi}{3}}|12\rangle +                             |21\rangle)_{ta}\\&
                      + \beta_2(e^{\text{i} \frac{2\pi}{3}}|02\rangle +                            |11\rangle +  e^{\text{i} \frac{4\pi}{3}}|20\rangle)_{ta}]
  + \alpha_1|1\rangle_c[\beta_0(                           |00\rangle + e^{\text{i} \frac{4\pi}{3}}|12\rangle +  e^{\text{i} \frac{2\pi}{3}}|21\rangle)_{ta} \\&
                      + \beta_1(e^{\text{i} \frac{4\pi}{3}}|02\rangle + e^{\text{i} \frac{2\pi}{3}}|11\rangle +                             |20\rangle)_{ta}
                      + \beta_2(e^{\text{i} \frac{2\pi}{3}}|01\rangle +                            |10\rangle +  e^{\text{i} \frac{4\pi}{3}}|22\rangle)_{ta}]\\&
  + \alpha_2|2\rangle_c[\beta_0(                           |02\rangle + e^{\text{i} \frac{4\pi}{3}}|11\rangle +  e^{\text{i} \frac{2\pi}{3}}|20\rangle)_{ta} 
                      + \beta_1(e^{\text{i} \frac{4\pi}{3}}|01\rangle + e^{\text{i} \frac{2\pi}{3}}|10\rangle +                             |22\rangle)_{ta}\\&
                      + \beta_2(e^{\text{i} \frac{2\pi}{3}}|00\rangle +                            |12\rangle +  e^{\text{i} \frac{4\pi}{3}}|21\rangle)_{ta}].
\end{split}
\end{equation}

If $\mathcal{P}_2=2$, then state $|\psi_0\rangle$ becomes
\begin{equation} \label{eq16}
\begin{split}
|\psi_{02}\rangle =&\ 
    \alpha_0|0\rangle_c[\beta_0(                           |00\rangle + e^{\text{i} \frac{2\pi}{3}}|12\rangle  +  e^{\text{i} \frac{4\pi}{3}}|21\rangle)_{ta} 
                      + \beta_1(e^{\text{i} \frac{2\pi}{3}}|02\rangle + e^{\text{i} \frac{4\pi}{3}}|11\rangle  +                             |20\rangle)_{ta}\\&
                      + \beta_2(e^{\text{i} \frac{4\pi}{3}}|01\rangle +                            |10\rangle  + e^{\text{i} \frac{2\pi}{3}} |22\rangle)_{ta}]
  + \alpha_1|1\rangle_c[\beta_0(                           |02\rangle + e^{\text{i} \frac{2\pi}{3}}|11\rangle  + e^{\text{i} \frac{4\pi}{3}} |20\rangle)_{ta} \\&
                      + \beta_1(e^{\text{i} \frac{2\pi}{3}}|01\rangle + e^{\text{i} \frac{4\pi}{3}}|10\rangle  +                             |22\rangle)_{ta}
                      + \beta_2(e^{\text{i} \frac{4\pi}{3}}|00\rangle +                            |12\rangle  + e^{\text{i} \frac{2\pi}{3}} |21\rangle)_{ta}]\\&
  + \alpha_2|2\rangle_c[\beta_0(                           |01\rangle + e^{\text{i} \frac{2\pi}{3}}|10\rangle  + e^{\text{i} \frac{4\pi}{3}} |22\rangle)_{ta} 
                      + \beta_1(e^{\text{i} \frac{2\pi}{3}}|00\rangle + e^{\text{i} \frac{4\pi}{3}}|12\rangle  +                             |21\rangle)_{ta}\\&
                      + \beta_2(e^{\text{i} \frac{4\pi}{3}}|02\rangle +                            |11\rangle  + e^{\text{i} \frac{2\pi}{3}} |20\rangle)_{ta}];
\end{split}
\end{equation}
State $|\psi_1\rangle$ becomes
\begin{equation} \label{eq17}
\begin{split}
|\psi_{12}\rangle =&\ 
    \alpha_0|0\rangle_c[\beta_0(                           |02\rangle + e^{\text{i} \frac{2\pi}{3}}|11\rangle  +  e^{\text{i} \frac{4\pi}{3}}|20\rangle)_{ta} 
                      + \beta_1(e^{\text{i} \frac{2\pi}{3}}|01\rangle + e^{\text{i} \frac{4\pi}{3}}|10\rangle  +                             |22\rangle)_{ta}\\&
                      + \beta_2(e^{\text{i} \frac{4\pi}{3}}|00\rangle +                            |12\rangle  +  e^{\text{i} \frac{2\pi}{3}}|21\rangle)_{ta}]
  + \alpha_1|1\rangle_c[\beta_0(                           |01\rangle + e^{\text{i} \frac{2\pi}{3}}|10\rangle  +  e^{\text{i} \frac{4\pi}{3}}|22\rangle)_{ta} \\&
                      + \beta_1(e^{\text{i} \frac{2\pi}{3}}|00\rangle + e^{\text{i} \frac{4\pi}{3}}|12\rangle  +                             |21\rangle)_{ta}
                      + \beta_2(e^{\text{i} \frac{4\pi}{3}}|02\rangle +                            |11\rangle  +  e^{\text{i} \frac{2\pi}{3}}|20\rangle)_{ta}]\\&
  + \alpha_2|2\rangle_c[\beta_0(                           |00\rangle + e^{\text{i} \frac{2\pi}{3}}|12\rangle  +  e^{\text{i} \frac{4\pi}{3}}|21\rangle)_{ta} 
                      + \beta_1(e^{\text{i} \frac{2\pi}{3}}|02\rangle + e^{\text{i} \frac{4\pi}{3}}|11\rangle  +                             |20\rangle)_{ta}\\&
                      + \beta_2(e^{\text{i} \frac{4\pi}{3}}|01\rangle +                            |10\rangle  +  e^{\text{i} \frac{2\pi}{3}}|22\rangle)_{ta}];
\end{split}
\end{equation}
State $|\psi_2\rangle$ becomes
\begin{equation} \label{eq18}
\begin{split}
|\psi_{22}\rangle =&\ 
    \alpha_0|0\rangle_c[\beta_0(                           |01\rangle + e^{\text{i} \frac{2\pi}{3}}|10\rangle  + e^{\text{i} \frac{4\pi}{3}}|22\rangle)_{ta} 
                      + \beta_1(e^{\text{i} \frac{2\pi}{3}}|00\rangle + e^{\text{i} \frac{4\pi}{3}}|12\rangle  +                            |21\rangle)_{ta}\\&
                      + \beta_2(e^{\text{i} \frac{4\pi}{3}}|02\rangle +                            |11\rangle  + e^{\text{i} \frac{2\pi}{3}}|20\rangle)_{ta}]
  + \alpha_1|1\rangle_c[\beta_0(                           |00\rangle + e^{\text{i} \frac{2\pi}{3}}|12\rangle  + e^{\text{i} \frac{4\pi}{3}}|21\rangle)_{ta} \\&
                      + \beta_1(e^{\text{i} \frac{2\pi}{3}}|02\rangle + e^{\text{i} \frac{4\pi}{3}}|11\rangle  +                            |20\rangle)_{ta}
                      + \beta_2(e^{\text{i} \frac{4\pi}{3}}|01\rangle +                            |10\rangle  + e^{\text{i} \frac{2\pi}{3}}|22\rangle)_{ta}]\\&
  + \alpha_2|2\rangle_c[\beta_0(                           |02\rangle + e^{\text{i} \frac{2\pi}{3}}|11\rangle  + e^{\text{i} \frac{4\pi}{3}}|20\rangle)_{ta} 
                      + \beta_1(e^{\text{i} \frac{2\pi}{3}}|01\rangle + e^{\text{i} \frac{4\pi}{3}}|10\rangle  +                            |22\rangle)_{ta}\\&
                      + \beta_2(e^{\text{i} \frac{4\pi}{3}}|00\rangle +                             |12\rangle  + e^{\text{i} \frac{2\pi}{3}}|21\rangle)_{ta}].
\end{split}
\end{equation}

In the third step, the output ancilla qutrit is measured in the basis $\{|0\rangle, |1\rangle, |2\rangle\}$. 
Then, based on the outcomes of $P_1$ and $P_2$, as well as the measurement result of ancilla qutrit, classical feed-forward operations are performed on the control and target qutrits to complete the generalized CNOT gate (see Table \ref{Tab2}). 
That is, the above operations transform the states $|\psi_{00}\rangle$, $|\psi_{01}\rangle$,     $|\psi_{02}\rangle$, 
                                                   $|\psi_{10}\rangle$, $|\psi_{11}\rangle$,     $|\psi_{12}\rangle$,
                                                   $|\psi_{20}\rangle$, $|\psi_{21}\rangle$, and $|\psi_{22}\rangle$ 
into the desired state
\begin{equation} \label{eq19}
\begin{split}
|\psi_{\text{out}}\rangle =&\ \alpha_0|0\rangle_c (\beta_0|0\rangle_t + \beta_1|1\rangle_t + \beta_2|2\rangle_t) 
                            + \alpha_1|1\rangle_c (\beta_0|1\rangle_t + \beta_1|2\rangle_t + \beta_2|0\rangle_t) \\& 
                            + \alpha_2|2\rangle_c (\beta_0|2\rangle_t + \beta_1|0\rangle_t + \beta_2|1\rangle_t).
\end{split}
\end{equation}

\begin{table}[htbp]
\centering 
\caption{The correspondences between the outputs $\mathcal{P}_1$, $\mathcal{P}_2$, and the ancilla qutrit in the construction of a qutrit-based generalized CNOT gate. 
Single-qutrit operation $R_{\lambda,3}^{(st)}(\theta)$ is given in Eqs. \eqref{eq20}-\eqref{eq23}.}
\label{Tab2}
\begin{tabular}{cccccc}
\hline\hline
\multirow{2}*{$\mathcal{P}_1$} \quad & \multirow{2}*{Ancilla} \quad & \multicolumn{3}{c}{Control qutrit} & \multirow{2}*{Target qutrit} \\
\cline{3-5}
& & $\mathcal{P}_2 = 0$ & $\mathcal{P}_2 = 1$ & $\mathcal{P}_2 = 2$ & \\
\hline
\multirow{3}{*}{0} 
  & $|0\rangle_a$ & $I_{3}$ & $R_{z,3}^{(12)}(2\pi/3)$  & $R_{z,3}^{(12)}(-2\pi/3)$ & $R_{x,3}^{(12)}(\pi/2)$ \\
  & $|1\rangle_a$ & $I_{3}$ & $R_{z,3}^{(01)}(2\pi/3)$  & $R_{z,3}^{(01)}(-2\pi/3)$ & $R_{x,3}^{(02)}(\pi/2)$ \\
  & $|2\rangle_a$ & $I_{3}$ & $R_{z,3}^{(02)}(-2\pi/3)$ & $R_{z,3}^{(02)}(2\pi/3)$  & $R_{x,3}^{(01)}(\pi/2)$ \\
\hline
\multirow{3}{*}{1} 
  & $|0\rangle_a$ & $I_{3}$ & $R_{z,3}^{(01)}(2\pi/3)$  & $R_{z,3}^{(01)}(-2\pi/3)$ & $R_{x,3}^{(02)}(\pi/2)$ \\
  & $|1\rangle_a$ & $I_{3}$ & $R_{z,3}^{(02)}(-2\pi/3)$ & $R_{z,3}^{(02)}(2\pi/3)$  & $R_{x,3}^{(01)}(\pi/2)$ \\
  & $|2\rangle_a$ & $I_{3}$ & $R_{z,3}^{(12)}(2\pi/3)$  & $R_{z,3}^{(12)}(-2\pi/3)$ & $R_{x,3}^{(12)}(\pi/2)$ \\
\hline
\multirow{3}{*}{2} 
  & $|0\rangle_a$ & $I_{3}$ & $R_{z,3}^{(02)}(-2\pi/3)$ & $R_{z,3}^{(02)}(2\pi/3)$  & $R_{x,3}^{(01)}(\pi/2)$ \\
  & $|1\rangle_a$ & $I_{3}$ & $R_{z,3}^{(12)}(2\pi/3)$  & $R_{z,3}^{(12)}(-2\pi/3)$ & $R_{x,3}^{(12)}(\pi/2)$ \\
  & $|2\rangle_a$ & $I_{3}$ & $R_{z,3}^{(01)}(2\pi/3)$  & $R_{z,3}^{(01)}(-2\pi/3)$ & $R_{x,3}^{(02)}(\pi/2)$ \\
\hline\hline
\end{tabular}
\end{table}
In Table \ref{Tab2}, the single-qutrit operation $R_{\lambda,3}^{(st)}(\theta)$ is given by
\begin{equation} \label{eq20}
  R_{\lambda,3}^{(st)}(\theta) = e^{-\text{i} \theta \sigma_{\lambda,3}^{(st)}},
\end{equation}
where $\sigma_{\lambda,3}^{(st)}$ denotes the generator of the Lie algebra $\text{su}(3)$, given by
\begin{equation}\label{eq21}
 \sigma_{z,d}^{(st)}=|s\rangle\langle s|-|t\rangle\langle t|,
\end{equation}
\begin{equation}\label{eq22}
 \sigma_{x,d}^{(st)}=|s\rangle\langle t|+|t\rangle\langle s|,
\end{equation}
\begin{equation}\label{eq23}
 \sigma_{y,d}^{(st)}=-\textrm{i}|s\rangle\langle t|+\textrm{i}|t\rangle\langle s|.
\end{equation}
Here, $s$ and $t$ are integers, and $0\leq s<t\leq 2$.

\section{Module-based two-qudit generalized CNOT gate} \label{Sec3}

The procedure presented in Sec. \ref{Sec2} can be extended to construct a qudit-based ($d \geq 4$) generalized CNOT gate. 
We can follow the same steps of that for the two-qutrit CNOT gate shown in Fig. \ref{Fig1}. 
We should only increase the dimension of each subsystem. 

(1) The input states of the control, target, and ancilla qutrits should be extended to the qudit-based scenario, i.e.,
\begin{equation}\label{eq24}
  |x\rangle_c = \alpha_{0}|0\rangle_c + \alpha_{1}|1\rangle_c + \cdots + \alpha_{d-1}|d-1\rangle_c,
\end{equation}
\begin{equation}\label{eq25}
  |y\rangle_t = \beta_{0} |0\rangle_t + \beta_{1} |1\rangle_t + \cdots + \beta_{d-1} |d-1\rangle_t,
\end{equation}
\begin{equation}\label{eq26}
  |z\rangle_a = \frac{1}{\sqrt{d}}(|0\rangle_a +|1\rangle_a + \cdots +|d-1\rangle_a).
\end{equation}

(2) The qutrit-based generalized parity module $\mathcal{P} =(j \ominus i) \bmod 3$ (depicted in Table \ref{Tab1}) should be extended to the qudit-based scenario $\mathcal{P} =(j \ominus i) \bmod d$ (depicted in Table \ref{Tab3}). 

\begin{table} [htbp]
\centering
\caption{The correspondences between the basis and the outputs in qudit-based generalized parity module with $\mathcal{P} =(j \ominus i) \bmod d$.}
\label{Tab3}

\begin{tabular}{ccccccc}
\hline\hline

      \multicolumn {4}{c}{Basis}                                            &  Output $\mathcal{P}$ & Class\\

\hline

      $|00\rangle$      & $|11\rangle$  & $\cdots$  & $|(d-1)(d-1)\rangle$  &  0                   & \uppercase\expandafter{\romannumeral1} \\

      $|01\rangle$      & $|12\rangle$  & $\cdots$  & $|(d-1)(d-d)\rangle$  &  1                   & \uppercase\expandafter{\romannumeral2} \\

      $\vdots$          & $\vdots$      & $\vdots$  & $\vdots$              &  $\vdots$            & $\vdots$ \\

      $|0(d-1)\rangle$  & $|10\rangle$  & $\cdots$  & $|(d-1)(d-2)\rangle$  &   $d-1$                & \uppercase\expandafter{\romannumeral1000} \\

\hline \hline
\end{tabular}
\end{table}

(3) Qutrit-based Fourier transformation $F_3$ described in Eq. \eqref{eq9} should be extended to the qudit-based scenario, i.e.,
\begin{equation}\label{eq27}
  F_{d}|m\rangle=\frac{1}{\sqrt{d}} \sum_{n=0}^{d-1} e^{\mathrm{i} 2\pi m n / d} |n\rangle.
\end{equation}
Here, $ m=0,1,\cdots, (d-1)$.

(4) The generators of Lie algebra of $\text{su}(3)$ given in Eqs. \eqref{eq21}-\eqref{eq23} can be generalized to the $\text{su}(d)$ scenario, that is, $0\leq s<t\leq (d-1)$.

(5) The classical feed-forward operations shown in Table \ref{Tab2} should be extended to qudit-based scenario.

The transformation of control qudit induced by the classical feed-forward operations is given by
\begin{equation}\label{eq28}
  |\varsigma\rangle_c \rightarrow \omega^{[-\mathcal{P}_2 \cdot (\varsigma + k + \mathcal{P}_1)]\bmod d} |\varsigma\rangle_c,
\end{equation}
where $\omega = e^{\text{i} 2\pi / d}$.
That is, this feed-forward operation  is governed by 
the outputs $\mathcal{P}_1$, $\mathcal{P}_2$, and the outcome of the ancilla qudit $|k\rangle_a$. Here, $\mathcal{P}_1, \mathcal{P}_2, k, \varsigma = 0,1,\cdots,(d-1)$.

The transformation of the target qudit induced by the classical feed-forward operations is given by
\begin{equation}\label{eq29}
  |\xi\rangle_t \rightarrow |-(\mathcal{P}_1 + \xi + k) \bmod d\rangle_t.
\end{equation}
That is, this feed-forward operation is governed by the output $\mathcal{P}_1$ and the outcome of the ancilla qudit $|k\rangle_a$. Here, $\mathcal{P}_1, k, \xi=0,1,\cdots,(d-1)$.

To clarify our description, we will take $d=4$, $\mathcal{P}_1 = 1$, $\mathcal{P}_2 = 1$, and $k=0$ as an example. 
Based on Eq. \eqref{eq28}, the classical feed-forward operation makes the control ququart become 
\begin{eqnarray}\label{eq_ex1}
  &&|0\rangle_c \rightarrow \omega^{[-1 \cdot (0 + 0 + 1)]\bmod 4}|0\rangle_c = \omega^{3}|0\rangle_c = -\text{i}|0\rangle_c,\\
  &&|1\rangle_c \rightarrow \omega^{[-1 \cdot (1 + 0 + 1)]\bmod 4}|1\rangle_c = \omega^{2}|1\rangle_c =         -|1\rangle_c,\\
  &&|2\rangle_c \rightarrow \omega^{[-1 \cdot (2 + 0 + 1)]\bmod 4}|2\rangle_c = \omega^{1}|2\rangle_c = \text{i} |2\rangle_c,\\
  &&|3\rangle_c \rightarrow \omega^{[-1 \cdot (3 + 0 + 1)]\bmod 4}|3\rangle_c = \omega^{0}|3\rangle_c =          |3\rangle_c.
\end{eqnarray}
Based on Eq. \eqref{eq29}, the classical feed-forward operation makes the target ququart become
\begin{eqnarray}\label{eq_ex2}
  &&|0\rangle_t \rightarrow |-(1 + 0 + 0) \bmod 4\rangle_t = |-1 \bmod 4\rangle_t = |3\rangle_t,\\
  &&|1\rangle_t \rightarrow |-(1 + 1 + 0) \bmod 4\rangle_t = |-2 \bmod 4\rangle_t = |2\rangle_t,\\
  &&|2\rangle_t \rightarrow |-(1 + 2 + 0) \bmod 4\rangle_t = |-3 \bmod 4\rangle_t = |1\rangle_t,\\
  &&|3\rangle_t \rightarrow |-(1 + 3 + 0) \bmod 4\rangle_t = |-4 \bmod 4\rangle_t = |0\rangle_t.
\end{eqnarray}


\section{Optical architecture for implementing generalized parity module} \label{Sec4}

\begin{figure}[htpb]
\begin{center}
\includegraphics[width=13.5 cm]{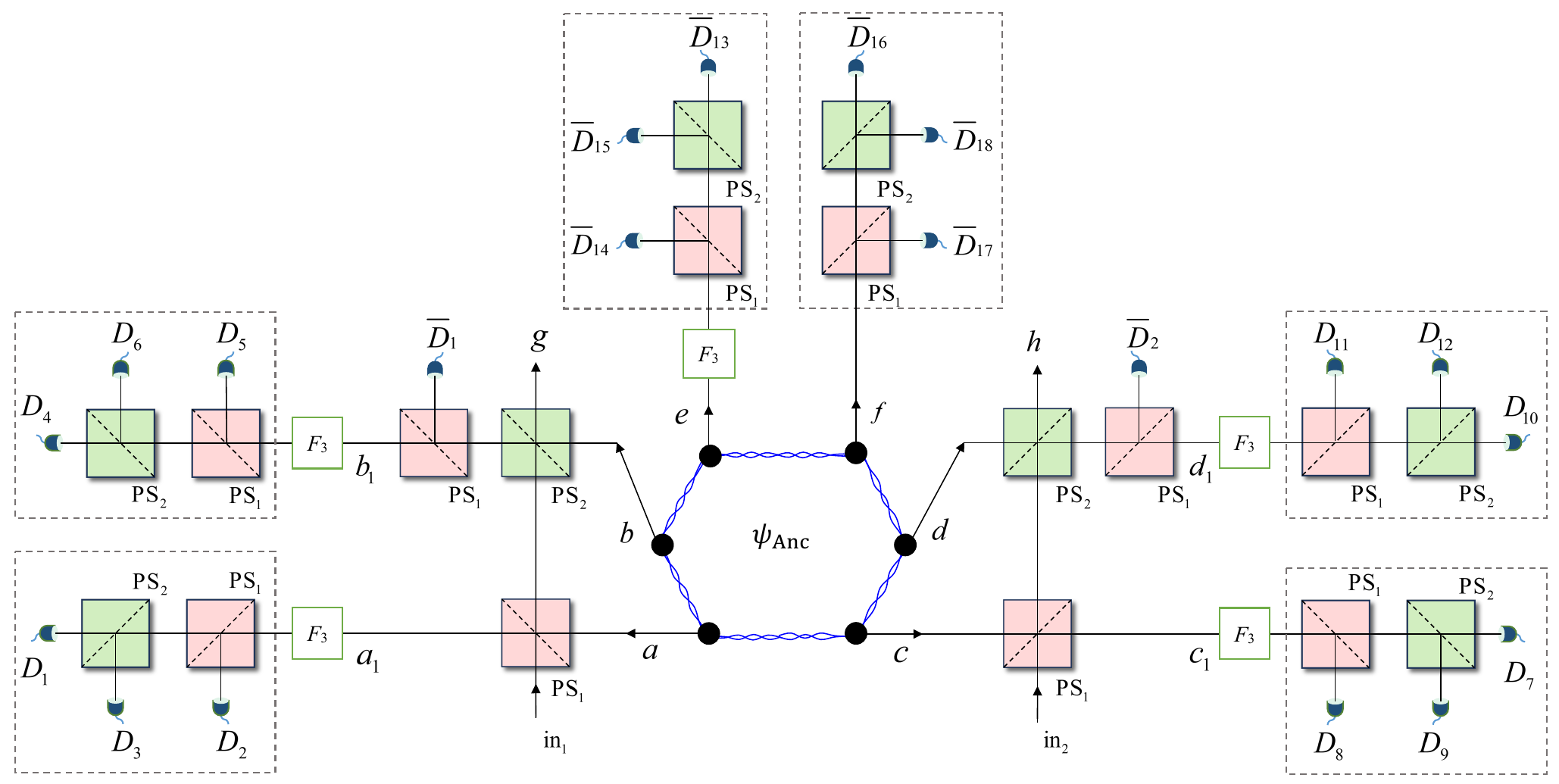}
\caption{Schematic diagram of generalized qutrit-based parity module. 
         $D$ and $\overline{D}$ denote photon-number-resolving detector and single-photon detector, respectively.         
         $F_3$ is qutrit-based Fourier transformation.
         The circuits in the dashed boxes are designed to measure the output photon in the basis $\{|0\rangle, |1\rangle,|2\rangle\}$.}
\label{Fig2}
\end{center}
\end{figure}

The key ingredient of the proposed scheme for implementing generalized CNOT gate is the high-dimensional generalized parity module. 
The optical architecture we designed for implementing the heralded generalized qutrit-based parity module is shown in Fig. \ref{Fig2}.

As depicted in Fig. \ref{Fig2}, the qutrits are encoded in the OAM DOF of single-photon systems, i.e., 
\begin{equation}\label{eq30}
 |0\rangle \equiv |\ell = 0\rangle,\;\;
 |1\rangle \equiv |\ell = 1\rangle,\;\;
 |2\rangle \equiv |\ell = 2\rangle.
\end{equation}
Without loss of generality, we consider the two input gate photons in the following arbitrary normalized states
\begin{equation}\label{eq31}
 |\phi\rangle_{\text{in}_1} = \gamma_0 |0\rangle + \gamma_1 |1\rangle + \gamma_2 |2\rangle,\;\;
 |\phi\rangle_{\text{in}_2} = \delta_0 |0\rangle + \delta_1 |1\rangle + \delta_2 |2\rangle.
\end{equation}
%
%
The six-photon ancillary state is prepared as
\begin{equation}\label{eq32}
\begin{split}
  |\psi_{\text{Anc}}\rangle=&\ \frac{1}{3}(|00\rangle(|00\rangle|00\rangle+|10\rangle|11\rangle+|02\rangle|22\rangle)
                                  + |10\rangle(|00\rangle|12\rangle+|10\rangle|20\rangle\\
                              & + |02\rangle|01\rangle) + |02\rangle(|00\rangle|21\rangle+|10\rangle|02\rangle+|02\rangle|10\rangle)).
\end{split}
\end{equation}

First, as illustrated in Fig. \ref{Fig2}, the control and the target photons are injected into the input ports  $\text{in}_1$ and $\text{in}_2$, respectively.
The first (middle) two photons contained in $|\psi_{\text{Anc}}\rangle$ are injected into the spatial modes $a$ and $b$ ($c$ and $d$), respectively.
The last two photons contained in $|\psi_{\text{Anc}}\rangle$ are directly routed to $e$ and $f$, respectively.
Then, the initial state of the composite system $|\Pi_0\rangle = |\phi\rangle_{\text{in}_1} \otimes |\phi\rangle_{\text{in}_2} \otimes |\psi_{\text{Anc}}\rangle$ can be rewritten as
\begin{equation}\label{eq33}
\begin{split}
|\Pi_0\rangle=&\ \frac{1}{3}(\gamma_0 |0\rangle_{\text{in}_1} + \gamma_1 |1\rangle_{\text{in}_1} + \gamma_2 |2\rangle_{\text{in}_1})  \otimes  (\delta_0 |0\rangle_{\text{in}_2} + \delta_1 |1\rangle_{\text{in}_2} + \delta_2 |2\rangle_{\text{in}_2}) \\
                      & \otimes  (|0\rangle_a|0\rangle_b|0\rangle_c|0\rangle_d|0\rangle_e|0\rangle_f 
                  +|0\rangle_a|0\rangle_b|1\rangle_c|0\rangle_d|1\rangle_e|1\rangle_f\\&
                  +|0\rangle_a|0\rangle_b|0\rangle_c|2\rangle_d|2\rangle_e|2\rangle_f               
                  +|1\rangle_a|0\rangle_b|0\rangle_c|0\rangle_d|1\rangle_e|2\rangle_f\\&  
                  +|1\rangle_a|0\rangle_b|1\rangle_c|0\rangle_d|2\rangle_e|0\rangle_f
                  +|1\rangle_a|0\rangle_b|0\rangle_c|2\rangle_d|0\rangle_e|1\rangle_f\\&            
                  +|0\rangle_a|2\rangle_b|0\rangle_c|0\rangle_d|2\rangle_e|1\rangle_f
                  +|0\rangle_a|2\rangle_b|1\rangle_c|0\rangle_d|0\rangle_e|2\rangle_f\\&
                  +|0\rangle_a|2\rangle_b|0\rangle_c|2\rangle_d|1\rangle_e|0\rangle_f).
\end{split}
\end{equation}

Second, after the incident photons and ancillary photons interact with $\text{PS}_1$s and $\text{PS}_2$s, 
we choose the instances in which each of the spatial modes $a_1$, $b_1$, $c_1$, $d_1$ contains exactly one photon (which can be achieved by employing  photon-number-resolving detectors), and then state $|\Pi_0\rangle$ collapse to
\begin{equation}\label{eq34}
\begin{split}
|\Pi_1\rangle=&\ \frac{1}{3}[
                 \gamma_0\delta_0|0\rangle_{a_1}|0\rangle_{b_1}|0\rangle_{c_1}|0\rangle_{d_1}|0\rangle_e|0\rangle_f|0\rangle_g|0\rangle_{h}\\&
                +\gamma_0\delta_2|0\rangle_{a_1}|0\rangle_{b_1}|0\rangle_{c_1}|2\rangle_{d_1}|2\rangle_e|2\rangle_f|0\rangle_g|2\rangle_h\\&
                +\gamma_2\delta_0|0\rangle_{a_1}|2\rangle_{b_1}|0\rangle_{c_1}|0\rangle_{d_1}|2\rangle_e|1\rangle_f|2\rangle_g|0\rangle_h\\&
                +\gamma_2\delta_2|0\rangle_{a_1}|2\rangle_{b_1}|0\rangle_{c_1}|2\rangle_{d_1}|1\rangle_e|0\rangle_f|2\rangle_g|2\rangle_h\\&
                +\frac{1}{2}(\gamma_1\delta_1|1\rangle_{a_1}|0\rangle_{b_1}|1\rangle_{c_1}|0\rangle_{d_1}|2\rangle_e|0\rangle_f|1\rangle_g|1\rangle_h\\& 
                +\gamma_1\delta_1|1\rangle_{a_1}|0\rangle_{b_1}|1\rangle_{c_1}|0\rangle_{d_1}|2\rangle_e|0\rangle_f|1\rangle_g|1\rangle_{\overline{D}_{2}}\\&
                +\gamma_1\delta_1|1\rangle_{a_1}|0\rangle_{b_1}|1\rangle_{c_1}|0\rangle_{d_1}|2\rangle_e|0\rangle_f|1\rangle_{\overline{D}_{1}}|1\rangle_h\\&
                +\gamma_1\delta_1|1\rangle_{a_1}|0\rangle_{b_1}|1\rangle_{c_1}|0\rangle_{d_1}|2\rangle_e|0\rangle_f|1\rangle_{\overline{D}_{1}}|1\rangle_{\overline{D}_{2}})\\&
                +\frac{1}{\sqrt{2}}(\gamma_0\delta_1|0\rangle_{a_1}|0\rangle_{b_1}|1\rangle_{c_1}|0\rangle_{d_1}|1\rangle_e|1\rangle_f|0\rangle_g|1\rangle_h\\&
                +\gamma_0\delta_1|0\rangle_{a_1}|0\rangle_{b_1}|1\rangle_{c_1}|0\rangle_{d_1}|1\rangle_e|1\rangle_f|0\rangle_g|1\rangle_{\overline{D}_{2}}\\&              
                +\gamma_2\delta_1|0\rangle_{a_1}|2\rangle_{b_1}|1\rangle_{c_1}|0\rangle_{d_1}|0\rangle_e|2\rangle_f|2\rangle_g|1\rangle_h\\&
                +\gamma_2\delta_1|0\rangle_{a_1}|2\rangle_{b_1}|1\rangle_{c_1}|0\rangle_{d_1}|0\rangle_e|2\rangle_f|2\rangle_g|1\rangle_{\overline{D}_{2}}\\&
                +\gamma_1\delta_0|1\rangle_{a_1}|0\rangle_{b_1}|0\rangle_{c_1}|0\rangle_{d_1}|1\rangle_e|2\rangle_f|1\rangle_g|0\rangle_{h}\\&
                +\gamma_1\delta_0|1\rangle_{a_1}|0\rangle_{b_1}|0\rangle_{c_1}|0\rangle_{d_1}|1\rangle_e|2\rangle_f|1\rangle_{\overline{D}_{1}}|0\rangle_{h}\\&
                +\gamma_1\delta_2|1\rangle_{a_1}|0\rangle_{b_1}|0\rangle_{c_1}|2\rangle_{d_1}|0\rangle_e|1\rangle_f|1\rangle_g|2\rangle_h\\&
                +\gamma_1\delta_2|1\rangle_{a_1}|0\rangle_{b_1}|0\rangle_{c_1}|2\rangle_{d_1}|0\rangle_e|1\rangle_f|1\rangle_{\overline{D}_{1}}|2\rangle_h)].
\end{split}
\end{equation}
Here, the linear optical element PS$_1$, the first-order parity sorter \cite{leach2002measuring, Yasir2021}, transmits the photons with $\ell = 0, 2$, and reflects those with  $\ell=1, 3$.
Similarly, PS$_2$, the second-order parity sorter \cite{gao2020computer}, transmits the photons with $\ell= 0$, reflects those with $\ell= 2$, and randomly transmits or reflects those with $\ell= 1$.
The details of the above two ``sorter''s are provided in Sec \ref{Sec5}.

Based on Eq. \eqref{eq34}, one can find that the click of single-photon detector $\overline{D}_{1}$ or $\overline{D}_{2}$ indicates that the generalized parity gate operation is a failure. 
While if $\overline{D}_{1}$ and $\overline{D}_{2}$ are not fired, then the state $|\Pi_1\rangle$ collapses to
\begin{equation}\label{eq35}
\begin{split}
|\Pi_2\rangle=&\ \frac{1}{3}[
                 \gamma_0\delta_0|0\rangle_{a_1}|0\rangle_{b_1}|0\rangle_{c_1}|0\rangle_{d_1}|0\rangle_e|0\rangle_f|0\rangle_g|0\rangle_{h}\\&
                +\gamma_0\delta_2|0\rangle_{a_1}|0\rangle_{b_1}|0\rangle_{c_1}|2\rangle_{d_1}|2\rangle_e|2\rangle_f|0\rangle_g|2\rangle_h\\&
                +\gamma_2\delta_0|0\rangle_{a_1}|2\rangle_{b_1}|0\rangle_{c_1}|0\rangle_{d_1}|2\rangle_e|1\rangle_f|2\rangle_g|0\rangle_h\\&
                +\gamma_2\delta_2|0\rangle_{a_1}|2\rangle_{b_1}|0\rangle_{c_1}|2\rangle_{d_1}|1\rangle_e|0\rangle_f|2\rangle_g|2\rangle_h\\&
                +\frac{1}{2}        \gamma_1\delta_1|1\rangle_{a_1}|0\rangle_{b_1}|1\rangle_{c_1}|0\rangle_{d_1}|2\rangle_e|0\rangle_f|1\rangle_g|1\rangle_h\\&
                +\frac{1}{\sqrt{2}}(\gamma_0\delta_1|0\rangle_{a_1}|0\rangle_{b_1}|1\rangle_{c_1}|0\rangle_{d_1}|1\rangle_e|1\rangle_f|0\rangle_g|1\rangle_h\\&        
                +\gamma_2\delta_1|0\rangle_{a_1}|2\rangle_{b_1}|1\rangle_{c_1}|0\rangle_{d_1}|0\rangle_e|2\rangle_f|2\rangle_g|1\rangle_h\\&
                +\gamma_1\delta_0|1\rangle_{a_1}|0\rangle_{b_1}|0\rangle_{c_1}|0\rangle_{d_1}|1\rangle_e|2\rangle_f|1\rangle_g|0\rangle_h\\&
                +\gamma_1\delta_2|1\rangle_{a_1}|0\rangle_{b_1}|0\rangle_{c_1}|2\rangle_{d_1}|0\rangle_e|1\rangle_f|1\rangle_g|2\rangle_h)].
\end{split}
\end{equation}

Based on Eq. \eqref{eq35}, one can see that $\mathcal{P}$ can be heralded by the output state of the photons emitted from spatial mode $f$. 
$|0\rangle_f$  corresponds to the group $\{|00\rangle_{gh}, |11\rangle_{gh}, |22\rangle_{gh}\}$, i.e., $\mathcal{P} = 0$; 
$|1\rangle_f$  corresponds to the group $\{|01\rangle_{gh}, |12\rangle_{gh}, |20\rangle_{gh}\}$, i.e., $\mathcal{P} = 1$;
$|2\rangle_f$  corresponds to the group $\{|02\rangle_{gh}, |10\rangle_{gh}, |21\rangle_{gh}\}$, i.e., $\mathcal{P} = 2$.

Third, in order to complete the heralded “sorter”,  OAM-based Fourier transform $F_3$s are performed on the photons emitted from spatial modes $a_1$, $b_1$,$c_1$, $d_1$, and $e$. 
The operation $F_3$s transform $|\Pi_2\rangle$ into
\begin{equation}\label{eq36}
\begin{split}
|\Pi_3\rangle=&\ \frac{1}{27\sqrt{3}}[(\gamma_0\delta_0|0\rangle_{g}|0\rangle_{h}+\frac{1}{2}\gamma_1\delta_1|1\rangle_{g}|1\rangle_{h} + \gamma_2\delta_2|2\rangle_{g}|2\rangle_{h})\\&
                \otimes (|0\rangle_{D_1} |0\rangle_{D_4} |0\rangle_{D_7} |0\rangle_{D_{10}} |0\rangle_{\overline{D}_{13}} |0\rangle_{\overline{D}_{16}}
                + |0\rangle_{D_1} |0\rangle_{D_4} |1\rangle_{D_8} |0\rangle_{D_{10}} |0\rangle_{\overline{D}_{13}} |0\rangle_{\overline{D}_{16}}\\&
                + |0\rangle_{D_1} |0\rangle_{D_4} |2\rangle_{D_9} |0\rangle_{D_{10}} |0\rangle_{\overline{D}_{13}} |0\rangle_{\overline{D}_{16}}
                + |0\rangle_{D_1} |1\rangle_{D_5} |0\rangle_{D_7} |0\rangle_{D_{10}} |0\rangle_{\overline{D}_{13}} |0\rangle_{\overline{D}_{16}}\\&
                + |0\rangle_{D_1} |1\rangle_{D_5} |1\rangle_{D_8} |0\rangle_{D_{10}} |0\rangle_{\overline{D}_{13}} |0\rangle_{\overline{D}_{16}}
                + |0\rangle_{D_1} |1\rangle_{D_5} |2\rangle_{D_9} |0\rangle_{D_{10}} |0\rangle_{\overline{D}_{13}} |0\rangle_{\overline{D}_{16}}
                + \cdots)\\&
                +(\frac{1}{\sqrt{2}}\gamma_0\delta_1|0\rangle_{g}|1\rangle_{h}+\frac{1}{\sqrt{2}}\gamma_1\delta_2|1\rangle_{g}|2\rangle_{h} + \gamma_2\delta_0|2\rangle_{g}|0\rangle_{h})\\&
               \otimes(|0\rangle_{D_1} |0\rangle_{D_4} |0\rangle_{D_7} |0\rangle_{D_{10}} |0\rangle_{\overline{D}_{13}} |0\rangle_{\overline{D}_{17}}
                + |0\rangle_{D_1} |0\rangle_{D_4} |1\rangle_{D_8} |0\rangle_{D_{10}} |0\rangle_{\overline{D}_{13}} |0\rangle_{\overline{D}_{17}}\\&
                + |0\rangle_{D_1} |0\rangle_{D_4} |2\rangle_{D_9} |0\rangle_{D_{10}} |0\rangle_{\overline{D}_{13}} |0\rangle_{\overline{D}_{17}}
                + |0\rangle_{D_1} |1\rangle_{D_5} |0\rangle_{D_7} |0\rangle_{D_{10}} |0\rangle_{\overline{D}_{13}} |0\rangle_{\overline{D}_{17}}\\&
                + |0\rangle_{D_1} |1\rangle_{D_5} |1\rangle_{D_8} |0\rangle_{D_{10}} |0\rangle_{\overline{D}_{13}} |0\rangle_{\overline{D}_{17}}
                + |0\rangle_{D_1} |1\rangle_{D_5} |2\rangle_{D_9} |0\rangle_{D_{10}} |0\rangle_{\overline{D}_{13}} |0\rangle_{\overline{D}_{17}}
                + \cdots)\\&
                +(\gamma_0\delta_2|0\rangle_{g}|2\rangle_{h}+\frac{1}{\sqrt{2}}\gamma_1\delta_0|1\rangle_{g}|0\rangle_{h}+\frac{1}{\sqrt{2}}\gamma_2\delta_1|2\rangle_{g}|1\rangle_{h})\\&
               \otimes (|0\rangle_{D_1} |0\rangle_{D_4} |0\rangle_{D_7} |0\rangle_{D_{10}} |0\rangle_{\overline{D}_{13}} |0\rangle_{\overline{D}_{18}}
                + |0\rangle_{D_1} |0\rangle_{D_4} |1\rangle_{D_8} |0\rangle_{D_{10}} |0\rangle_{\overline{D}_{13}} |0\rangle_{\overline{D}_{18}}\\&
                + |0\rangle_{D_1} |0\rangle_{D_4} |2\rangle_{D_9} |0\rangle_{D_{10}} |0\rangle_{\overline{D}_{13}} |0\rangle_{\overline{D}_{18}}
                + |0\rangle_{D_1} |1\rangle_{D_5} |0\rangle_{D_7} |0\rangle_{D_{10}} |0\rangle_{\overline{D}_{13}} |0\rangle_{\overline{D}_{18}}\\&
                + |0\rangle_{D_1} |1\rangle_{D_5} |1\rangle_{D_8} |0\rangle_{D_{10}} |0\rangle_{\overline{D}_{13}} |0\rangle_{\overline{D}_{18}}
                + |0\rangle_{D_1} |1\rangle_{D_5} |2\rangle_{D_9} |0\rangle_{D_{10}} |0\rangle_{\overline{D}_{13}} |0\rangle_{\overline{D}_{18}}
                + \cdots)].
\end{split}
\end{equation}

Based on Eq. \eqref{eq36}, one can see that based on the clicks of the detectors, after some classical feed-forward operations are performed on the incident photons, the generalized parity module can be accomplished. 
The state in each group can be balanced by employing unbalanced beam splitters.

Therefore, the setup shown in Fig. \ref{Fig2} performs a heralded and nondestructive generalized qutrit-based parity module operation.

\section{Discussions and conclusion} \label{Sec5}

It has been demonstrated that parity gates supplemented with single-partite gates are sufficient for quantum computation, and the works mainly focused on qubit scenarios, see Tab \ref{Tab4}. 
In this paper, we develop a versatile generalized parity module $ \mathcal{P} =(j \ominus i) \bmod d$. 
Furthermore, we show that alternative form  of ``sorter'' may also be universal for quantum computing.
To clarify the significance of our work, a detailed comparison between our scheme and representative previous protocols is summarized in Table \ref{Tab4}.

\begin{table} [htbp]
\centering
\caption{Comparison of different parity-based CNOT schemes.}
\label{Tab4}
\begin{tabular}{cccccc}
\hline\hline
     Scheme                        &  $d$-level & System          & Mediate     & Detection               & Sorters \\
\hline
      Ref. \cite{CWJ2004charge}    &  $d = 2$   & Free electrons  & None        & Charge detection        & 1 \\
      Ref. \cite{Nemoto2004nearly} &  $d = 2$   & Photons         & Cross-Kerr  & Quadrature measurement  & 1 \\
      Our                          &  $d \ge 3$ & Photons         & None        & Polarization detection  & 2  \\
\hline \hline
\end{tabular}
\end{table}

\subsection{Alternative generalized parity module}

For a qudit system, there are $(d-1)!$ types of ``sorters''; 
crucially, we find that apart from $\mathcal{P} = (j \ominus i) \bmod d$, only sorter $\mathcal{P} = (j \oplus i) \bmod d$ is universal for quantum computing. 
The action of this non-trivial sorter is given by
\begin{equation} \label{eq37}
  |i j \rangle |0\rangle \rightarrow |i j\rangle |(j \oplus i) \bmod d\rangle.
\end{equation}
The correspondences between the input basis $|ij\rangle$ and the output $\mathcal{P}=(j \oplus i) \bmod d$ are shown in Table \ref{Tab5}. 
Based on $\mathcal{P}=(j \oplus i) \bmod d$, we can follow the same setup shown in Fig. \ref{Fig1} to construct a generalized CNOT gate, where only the classical feed-forward operations need to be modified.

\begin{table} [htbp]
\centering
\caption{Alternative qudit-based generalized parity module with $\mathcal{P} = (i \oplus j) \bmod d$.}
\label{Tab5}
\begin{tabular}{cccccc}
\hline\hline
      \multicolumn{4}{c}{Basis} & $\mathcal{P}$ & Class \\
\hline
      $|00\rangle$       & $|1(d-1)\rangle$  & $\cdots$   & $|(d-1)1\rangle$     &  0         & \text{I} \\
      $|01\rangle$       & $|10\rangle$      & $\cdots$   & $|(d-1)2\rangle$     &  1         & \text{II} \\
      $|02\rangle$       & $|11\rangle$      & $\cdots$   & $|(d-1)3\rangle$     &  2         & \text{III} \\
      $\vdots$           & $\vdots$          & $\vdots$   & $\vdots$             &  $\vdots$  & $\vdots$ \\
      $|0(d-2)\rangle$   & $|1(d-3)\rangle$  & $\cdots$   & $|(d-1)(d-1)\rangle$ &   $d-2$    & \text{M}-1 \\
      $|0(d-1)\rangle$   & $|1(d-2)\rangle$  & $\cdots$   & $|(d-1)0\rangle$     &   $d-1$    & \text{M} \\
\hline \hline
\end{tabular}
\end{table}

By evaluating their algebraic structures, we find that there are $(d-1)!$ possible ``sorters'', which correspond to the permutations of the second qudit index $i$ denoted by $\pi_m(i)$ with  $\pi_m(0)=0$. 
Specifically, the constraint $\pi_m(0)=0$ ensures that the first input port (reference state $|0\rangle$) remains invariant under the permutation. For a $d$-level system, this leaves $(d-1)!$ possible permutations of the remaining state indices $\{1, 2, \dots, d-1\}$. 
In the case of $d=4$, there are $3!$ permutations, which we index from $\pi_1$ to $\pi_6$ in lexicographical order, where $\pi_1 = (0, 1, 2, 3)$ represents the identity mapping and $\pi_2 = (0, 3, 2, 1)$ represents the fully reversed mapping.
It is worth pointing out that two of these ``sorters'', namely $\mathcal{P} = (j \ominus i) \bmod d$ (associated with $\pi_1$) and $\mathcal{P} = (j \oplus i) \bmod d$ (associated with $\pi_2$), can be recognized as building blocks for quantum computing. However, this does not mean that the other $(d-1)!-2$ non-linear permutation ``sorters'' are useless; they might be used for certain quantum information processing tasks, such as generating entangled states. Table \ref{Tab6} shows the six possible ``sorters'' for a 4-level system.

\begin{table} [htbp]
\centering
\caption{The 6 possible generalized 4-level parity module ``sorters''.}
\label{Tab6}
\begin{tabular}{ccccccc}
\hline\hline
Class & Sorter ($\pi_1$) & Sorter ($\pi_2$) & Sorter ($\pi_3$) & Sorter ($\pi_4$) & Sorter ($\pi_5$) & Sorter ($\pi_6$) \\
\hline
\multirow{2}{*}{I}          
      & $|00\rangle$, $|11\rangle$, & $|00\rangle$, $|13\rangle$, & $|00\rangle$, $|11\rangle$, & $|00\rangle$, $|12\rangle$, & $|00\rangle$, $|12\rangle$, & $|00\rangle$, $|13\rangle$, \\
      & $|22\rangle$, $|33\rangle$  & $|22\rangle$, $|31\rangle$  & $|23\rangle$, $|32\rangle$  & $|21\rangle$, $|33\rangle$  & $|23\rangle$, $|31\rangle$  & $|21\rangle$, $|32\rangle$  \\ \cline{2-7}
\multirow{2}{*}{II}          
      & $|01\rangle$, $|12\rangle$, & $|01\rangle$, $|10\rangle$, & $|01\rangle$, $|12\rangle$, & $|01\rangle$, $|13\rangle$, & $|01\rangle$, $|13\rangle$, & $|01\rangle$, $|10\rangle$, \\
      & $|23\rangle$, $|30\rangle$  & $|23\rangle$, $|32\rangle$  & $|20\rangle$, $|33\rangle$  & $|22\rangle$, $|30\rangle$  & $|20\rangle$, $|32\rangle$  & $|22\rangle$, $|33\rangle$  \\ \cline{2-7}
\multirow{2}{*}{III}        
      & $|02\rangle$, $|13\rangle$, & $|02\rangle$, $|11\rangle$, & $|02\rangle$, $|13\rangle$, & $|02\rangle$, $|10\rangle$, & $|02\rangle$, $|10\rangle$, & $|02\rangle$, $|11\rangle$, \\
      & $|20\rangle$, $|31\rangle$  & $|20\rangle$, $|33\rangle$  & $|21\rangle$, $|30\rangle$  & $|23\rangle$, $|31\rangle$  & $|21\rangle$, $|33\rangle$  & $|23\rangle$, $|30\rangle$  \\ \cline{2-7}
\multirow{2}{*}{IV}        
      & $|03\rangle$, $|10\rangle$, & $|03\rangle$, $|12\rangle$, & $|03\rangle$, $|10\rangle$, & $|03\rangle$, $|11\rangle$, & $|03\rangle$, $|11\rangle$, & $|03\rangle$, $|12\rangle$, \\
      & $|21\rangle$, $|32\rangle$  & $|21\rangle$, $|30\rangle$  & $|22\rangle$, $|31\rangle$  & $|20\rangle$, $|32\rangle$  & $|22\rangle$, $|30\rangle$  & $|20\rangle$, $|31\rangle$  \\
\hline \hline
\end{tabular}
\end{table}

\subsection{Evaluation of the optical generalized parity module}

The generalized parity module is a cornerstone for scalable high-dimensional quantum computing. 
Prior research has demonstrated that universal spatial-based single-qudit gates can be implemented with high fidelity using efficient interferometer decompositions \cite{interferometer1,interferometer2}. 
Reck \emph{et al.} \cite{interferometer1} and Clements \emph{et al.} \cite{interferometer2} showed that any discrete unitary operator, including single-qudit Fourier $F_d$, can be decomposed into an interferometric network consisting of beam splitters and phase shifters. 
Currently, such multiport interferometers can be reliably implemented on programmable universal linear optical platforms \cite{Carolan2015}.

As shown in Fig. \ref{Fig2}, ancillary state $|\psi_{\text{Anc}}\rangle$, shown in Eq. \eqref{eq32}, is critical for constructing generalized CNOT gates. 
Such high-dimensional states are prepared using ``Entanglement by Path Identity'' \cite{krenn2017entanglement} combined with multi-photon filters based on generalized Bell-state measurements \cite{high-BSA} to eliminate multi-photon noise without relying on post-selection.

As shown in Fig. \ref{Fig2}, parity sorters PS$_1$ and PS$_2$ are also key elements of our scheme. Generally, a parity sorter of order $N$ (where $N=1, 2$) consists of a symmetric Mach-Zehnder interferometer containing a Dove prism in each arm \cite{leach2002measuring,gao2020computer}. 
Due to the geometric rotation property of a Dove prism, orienting the two prisms at a relative angle of $\theta = \pi / (2N)$ rotates the transmitted beam profile by $2\theta = \pi/N$, which introduces an OAM-dependent phase shift
\begin{equation}\label{eq38}
    \Delta \phi = \ell \cdot (2\theta) = \frac{\ell\pi}{N},
\end{equation}
where $\ell$ represents the OAM topological charge. This phase difference enables the deterministic or probabilistic routing of different OAM modes.

\begin{figure}[!ht]
\begin{center}
\includegraphics[width=7.5 cm]{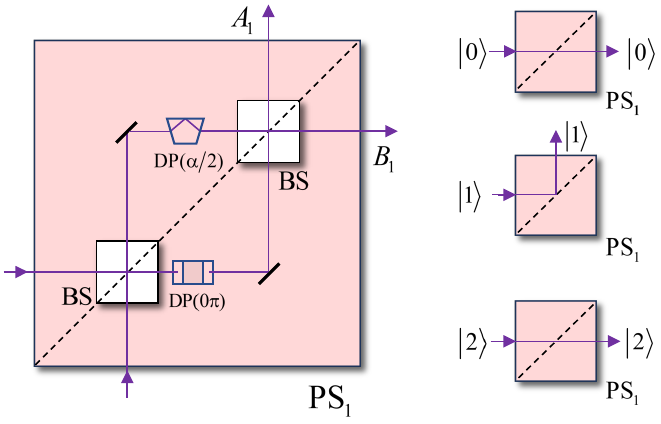}
\caption{Schematic diagram of a first-order parity sorter (PS$_{1}$) \cite{leach2002measuring}.
         Dove prism flips the transverse cross section of the beam in one arm while leaving the other unchanged.
         This Mach-Zehnder interferometer routes the photons with even $\ell$ ($l \bmod 2 = 0$) to Port $B_1$ (transmitted) and those with odd $\ell$ ($l \bmod 2 = 1$) to Port $A_1$ (reflected).} 
\label{Fig3}
\end{center}
\end{figure}

\emph{The first-order parity sorter PS$_1$.}   
As illustrated in Fig. \ref{Fig3}, first-order parity sorter PS$_1$ \cite{leach2002measuring} with $\alpha/2 = \pi/2$ (the relative angle between two Dove prisms) results in an effective beam rotation $\alpha = \pi$ of the transmitted beam.
By adjusting the path length of the interferometer, the device leads photons with even $\ell$ (i.e., $\ell \bmod 2 = 0$, e.g., $\ell = 0, \pm2, \pm4, \cdots $) to Port $B_1$ (transmitted) and those with odd $\ell$ (i.e., $\ell \bmod 2 = 1$, e.g., $\ell=\pm1, \pm3, \cdots $) to Port $A_1$ (reflected).

\begin{figure}[!ht]
\begin{center}
\includegraphics[width=8.0 cm]{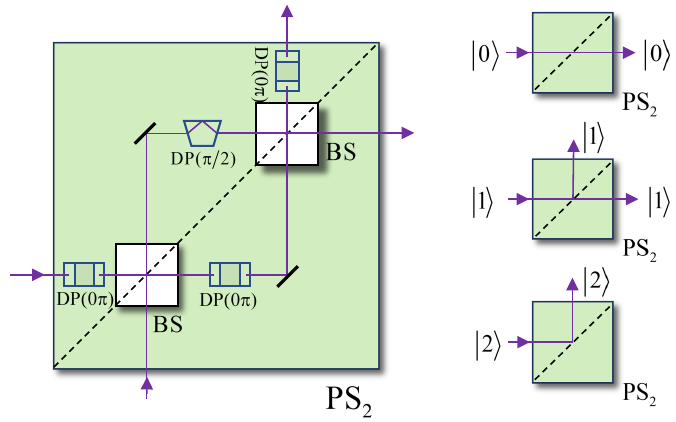}
\caption{Schematic diagram of a second-order parity sorter (PS$_2$) \cite{gao2020computer}. 
         This interferometric device classifies OAM modes by $\ell \bmod{4}$. 
         The device transmits photons with $\ell \bmod{4} = 0$ (e.g., $|0\rangle, |\pm 4\rangle$), and reflects the photons with $l \bmod{4} = 2$ (e.g., $|\pm 2\rangle, |\pm 6\rangle$). 
         For the photons with odd $\ell$ (e.g., $|\pm 1\rangle, |\pm 3\rangle$), they are randomly led to the two output ports.}
\label{Fig4}
\end{center}
\end{figure}

\emph{The second-order parity sorter PS$_2$}.
The architecture for realizing the second-order parity sorter (PS$_2$) \cite{gao2020computer} is shown in Fig. \ref{Fig4}.  
Such device transmits the components with $\ell \pmod 4 = 0$ (e.g., $|0\rangle, |\pm4\rangle, |\pm8\rangle, \cdots$),
             reflects the components with $\ell \pmod 4 = 2$ (e.g., $|\pm2\rangle, |\pm6\rangle, |\pm10\rangle, \cdots $), and
randomly directs the components with odd $\ell$ (e.g., $|\pm1\rangle, |\pm3\rangle, \cdots$) to the two output ports.

\smallskip

In practical applications, several factors contribute to reduce the performance of the
parity, including: 
The spatial-temporal mismatch between the incident photons and the six-photon ancilla state; 
Technical imperfections in the PS$_{1}$ and PS$_2$ sorters; 
The state purity of the ancillary source;  
Detector background/dark counts (typically 2\%).

In summary, we have proposed a scheme for constructing a module-based high-dimensional generalized CNOT gate. 
First, two alternative qutrit-based generalized parity modules are developed, and a program for constructing a qutrit-based generalized CNOT gate is proposed later. 
Subsequently, the procedures are expanded to arbitrary qudit-based scenarios.
The significant advantage of the proposed generalized CNOT scheme is that it is deterministic, heralded, and its performance is largely insensitive to the dimensionality of the computational qudits.
By encoding and decoding the quantum transformation in an ancillary mediator, the high-dimensional generalized parity module can be prepared via quantum nondemolition measurement, and individual ``sorters'' can be heralded by the ancillary mediator. 
Moreover, the proposed generalized parity module can also be used as building blocks for quantum communication.
Lastly, an optical architecture for implementing a qutrit-based generalized parity module is designed. 
The proposed architecture is nondemolition, and the success of this architecture is heralded by the detectors. 
The high-dimensional qutrits introduced here are encoded in the OAM DOFs of single-photon systems, but the concepts can be readily generalized to other high-dimensional encoding systems.
The evaluation of the parity gate confirms that our protocol is both theoretically sound and experimentally feasible.

\medskip

\section*{Funding} \par

This work was supported by the National Natural Science Foundation of China under Grant Nos. 62371038 and 12505028,
the Science Research Project of Hebei Education Department under Grant No. QN2025054, and the Beijing Natural Science Foundation under Grant No. 4252006.

\section*{Disclosures} \par
The authors declare no conflicts of interest.

\section*{Data availability} \par

The data supporting the conclusions of this study are available from the corresponding author upon reasonable request.

\end{document}